\documentclass[aps,pre,twocolumn,superscriptaddress]{revtex4-1}

\usepackage{graphicx}
\usepackage{todonotes}
\usepackage{hyperref}
\usepackage{amssymb}
\usepackage{mathrsfs}
\usepackage{amsmath}
\usepackage{verbatim}
\begin{document}


\title{Liquid-Liquid Critical Point in 3D Many-Body Water Model}


\author{Luis Enrique Coronas}
\affiliation{Facultat de F\'{\i}sica, Universitat de Barcelona, Diagonal
645, 08028 Barcelona, Spain.}

\author{Valentino Bianco}
\affiliation{Facultat de F\'{\i}sica, Universitat de Barcelona, Diagonal
645, 08028 Barcelona, Spain.}

\author{Arne Zantop}
\affiliation{Max Planck Institute for Dynamics and Self-Organization,
  Am Fassberg 17, 37077 G\"ottingen, Germany}

\author{Giancarlo Franzese}
\email{gfranzese@ub.edu}
\affiliation{Facultat de F\'{\i}sica, Universitat de Barcelona, Diagonal
645, 08028 Barcelona, Spain.}


\date{\today}

\begin{abstract}
Many-body interactions can play a relevant
role in water properties.
Here we study by Monte Carlo simulations  a coarse-grained
model for bulk water that includes many-body interactions associated to
water cooperativity. The model is efficient and allows us to
equilibrate water at extreme low temperatures in a wide range of pressures.
We find the line of temperatures of maximum density at constant
pressure and, at low temperature and high pressure, a phase transition
between high-density liquid and low-density liquid phases. The liquid-liquid
phase transition ends in a critical point. In the supercritical liquid region we
find for each thermodynamic response function  a locus of weak maxima at
high temperature and a locus of strong maxima at low temperature,
with both loci converging at the liquid-liquid critical point where
they diverge.
Comparing our results with previous works for the 
phase diagram of a many-body water monolayer, we observe that the
weak maxima of the specific heat are less evident in bulk and appear
only at negative 
pressures, while we find those 
of compressibility and thermal
expansion coefficient also at positive pressures in bulk.
However, the strong maxima of compressibility and thermal
expansion coefficient are very sharp for the bulk case.
Our results clarify fundamental properties of bulk water, possibly
difficult to detect with atomistic 
models not accounting for many-body interactions.
\end{abstract}


\maketitle

\section{Introduction}
Water models are of fundamental importance in many fields of physics,
chemistry, biology and engineering.
For example, it is now commonly accepted that water 
plays a key role in 
protein folding \cite{LevyPNAS2004, FranzeseFood2011,
  BiancoJBioPhys2012, BiancoPRL2015,folding2, Mallamace2016} and
protein dynamics 
\cite{Johnson:2008ys, Ngai:2008zr, crossover,dynamics2, fogarty2014,
  Mallamace2015}.
Computer simulations are an important 
tool to study such phenomena. 
Nevertheless, the computational cost for 
full-atom simulations of large-scale 
biological systems with explicit water  for experimentally-relevant time-scales 
is often prohibitive. 
Therefore, it is necessary to explore alternative approaches that
could allow us  to access mesoscopic scales and
time-scales of the order of seconds. A possible solution is to
develop  coarse-grain models of water that could 
greatly reduce  the simulation cost for the solvent preserving its
main physical properties.

Despite the importance of water in many aspects of life, there are still open questions concerning its complex
nature. Water exhibits more than 60 anomalies \cite{chaplin}, like 
the existence of a density maximum in the liquid phase at ambient 
pressure and temperature $T\sim 4^\circ$C 
or the anomalous increase of the specific heat $C_P$,
isothermal compressibility $K_T$, and absolute value of 
the thermal expansivity $\alpha_P$ upon cooling liquid water toward
the melting line and below it, in the supercooled liquid state \cite{debenedetti-book}.
A common understanding is that all these peculiar properties of water
are related to its hydrogen bonds (HBs)  forming 
a network that  orders following tetrahedral patterns at low $T$ \cite{Errington01}.
A large number of classical atomistic water
models have been proposed in the last decades with some performing
better than others \cite{comparison}. Nevertheless, there is still no
atomistic model able to fit all the water properties \cite{comparison}. 
One possible reason for it  is that pair-additive potentials do not
capture the effective many-body interactions due to 
quantum effects and polarizability \cite{quantum1, quantum2, quantum3,
  quantum4}. Consistent with this view, the effect of these many-body
interactions are stronger
where the water anomalies are more evident \cite{quantum5}.


The origin of anomalous properties of water has been largely debated
since the 80's \cite{kanno:4008, AngellC.A._j100395a032, poolePRL1994,
  Tanaka96, Mishima1998, SoperPRL2000,  Xu2005, LiuPNAS2007,
  StokelyPNAS2010, abascalJCP2010, bertrand2011, holtenJCP2013,
  pallares2014, BiancoSR2014, Soper2014, Nilsson2015,
  sciortinoPCCP2011, liuJCP2012, palmer_car_debenedetti, pooleJCP2013,
  Palmer2014, SmallenburgPRL2015} and a series of thermodynamic
scenarios have been proposed \cite{Speedy1982, Sastry1996, Poole92,
  AngellScience2008}. In particular,  Poole et al., 
based on molecular dynamic simulations, proposed a liquid-liquid
second critical point (LLCP) in the supercooled region
\cite{Poole92}. According to this scenario, the LLCP is located at the
end of  a first order phase transition separating two liquid
metastable phases of water with different density, structure and 
energy \cite{liuJCP2012,BiancoSR2014,Palmer2014}. 
Extrapolations from experimental data \cite{fuentevilla}, based on the hypothesis of such a
critical point, show that the maxima in response functions should be
much stronger and at lower temperature than those predicted by the classical model used in
Ref.~\cite{Poole92}. However, equilibrate 
classical models at very low $T$ is extremely computationally
demanding, making difficult to settle the  
disagreement between the simulations and the
extrapolated experimental data.


We propose a coarse grained (CG) bulk 
water model that includes many-body interactions in a
computationally effective way. Our results extend those for a
previous model of a many-body water monolayer, able to 
reproduces the main features of water in a
very efficient way and at those extreme
conditions that are not easily reachable by atomistic simulations
\cite{FMS2003, fs, widom, FS2007, delosSantos2011, BiancoSR2014}.
Here we show that the bulk
model exhibits a liquid-liquid phase transition (LLPT) in the supercooled
region, as the monolayer. We find a relevant difference between the
results for the response functions in the the bulk model and those in
the monolayer. In both cases we observe a line of weak maxima and a
line of strong maxima for the response functions in the
pressure-temperature phase diagram, however the $C_P$ weak maxima in
bulk water are observable only at negative  pressure, at variance with
the monolayer case.

The paper is organized as follows. In Section II we define the
coarse-grain model  for bulk water with effective many-body
interactions. In Section III we describe the simulation method. In
Section IV we present and discuss our results. In Section V we give
our conclusions. In Appendix A we present further details of the phase
diagram and in Appendix B we explain the details of the equilibration
and the error calculations.

\section{The Many-Body Water Model}
We consider a system of a constant number of molecules $N$, pressure
$P$ and temperature $T$ in a variable volume $V$.
We partition 
the total volume $V$ into a regular cubic lattice of ${\cal N}$ cells, $i
\in [1,...,{\cal N}]$, each with volume $v_i\equiv V/{\cal N}$.
For sake of simplicity we set ${\cal N}=N$ and consider the case of
a homogeneous fluid, in such a way that each cell is occupied 
by one water molecule. Next we  introduce a discretized  density
field $n_i \equiv \theta(2-v_i/v_0)$, where the Heaviside step
function $\theta(x)$ is 0 or 1 depending if $x<0$ or $\geq 0$,
corresponding to gas-like or liquid-like local density, respectively, 
and
$v_0\equiv 4/3 \pi (r_0/2)^3$ is the van der Waals volume  of a water
molecules, with $r_0 \equiv 2.9$~\AA.

The Hamiltonian of the system is by definition the sum of three
terms
\begin{equation}
\label{Ham}
\mathscr{H} \equiv \mathscr{H}_{\rm vdW} + \mathscr{H}_{\rm HB} +
\mathscr{H}_{\rm coop},
\end{equation}
corresponding to  the  van der Waals isotropic
interaction, the directional and the cooperative
contribution to  the HB interaction,  respectively.
A detailed description and motivation for each term can  
be found in Ref.s \cite{widom, delosSantos2011, BiancoSR2014}.

The van der Waals Term is given by the sum over the contributions of
all pairs of molecules $i$ and $j$,
\begin{equation}\label{Ham_vdW}
\mathscr{H}_{\rm vdW} \equiv \sum_{<i,j>} U(r_{ij}),
\end{equation}
where $U(r)$ is a Lennard-Jones potential with a hard-core (van der Waals)
diameter $r_0$, i.e.
\begin{equation}\label{Lennard_jones}
U(r) \equiv \left \lbrace
\begin{array}{ll}
\infty &  \textup{if} \ r \leq r_0 \\
\epsilon \left[ \left( \frac{r_0}{r} \right)^{12} - \left( \frac{r_0}{r} \right)^6 \right] & \textup{if} \  r_0 > r > r_c \\
0 & \textup{if} \  r_c < r.
\end{array}
\right.
\end{equation}
We truncate the interaction at a distance $r_c \equiv 6r_0$.

The second terms in Eq. (\ref{Ham}) is defined as
\begin{equation}\label{Ham_HB}
\mathscr{H}_{\rm HB} \equiv -J N_{\rm HB}\equiv -J \sum_{<ij>} n_in_j \delta_{ \sigma_{ij} \sigma_{ji} }
\end{equation}
where $N_{\rm HB}$ is the total number of HB in the system.
A water molecule  $i$ can form up to four HBs with molecules
$j$ at a distance such that $n_in_j=1$, i.e. $v_i\leq 2v_0$
implying a water-water distance $r_{i,j}\leq 2^{1/3} r_0 \simeq 3.7$~\AA.
Furthermore, to form a HB the relative orientation of the two water
molecules must be such that the angle between 
the OH of a molecules and the O-O direction does not exceed $\pm
30^o$. Hence, only 1/6 of the entire range of values $[0, 360^o]$
for this angle is associated to a bonding state.
We account for this constrain introducing a bonding variable 
$\sigma_{ij} \in [1,q]$ of molecule $i$ with each of the six nearest molecules $j$ and setting
$q= 6$ to account correctly for $i)$ the contribution to the HB energy via
the $ \delta_{ \sigma_{ij} \sigma_{ji} }\equiv 1$ if $ \sigma_{ij}=
\sigma_{ji}$, 0 otherwise, and $ii)$ the entropy variation due to the HB
formation and breaking.
We set $J / 4\epsilon = 0.5$, consistent with the proportion between the van der
Waals interaction and the directional (covalent) part of the HB interaction.

The last term in the Hamiltonian accounts for the cooperativity of the
HBs and is an effective many-body interaction between HBs
 due the O--O--O correlation that locally leads the molecules toward
 an ordered (tetrahedral) configuration
\begin{equation}\label{Ham_coop}
\mathscr{H}_{\rm coop} \equiv -J_{\sigma} N_{\rm coop} \equiv -J_{\sigma} \sum_{i} n_i \sum_{<k l>_i}\delta_{ \sigma_{ik} \sigma_{il}},
\end{equation}
where the sum is over the six bonding indices of the molecules $i$ with
the molecules $k$ and $l$, both near $i$.
By setting $J_\sigma / 4\epsilon = 0.03$, with $J_\sigma\ll J$, we guarantee that this term
is relevant only at those temperatures below which the molecule $i$
has already formed (non-tetrahedral)  HBs with the molecules $k$ and
$l$, implying an effective many-body interaction of molecule $i$ with
the four bounded molecules in its hydration shell.

A consequence of the formation of a network of tetrahedral HBs is the
increase of proper volume associated to each molecule.
The model takes this effect into account on average by adding a small increase of
volume $v_{\rm HB} = 0.5 v_0$ per bond, consistent with the average
volume increase between high-$\rho$ ices VI and VIII and low-$\rho$
(tetrahedral) ice Ih.
Therefore, the total volume of the
system with $N_{\rm HB}$ HBs is 
\begin{equation}\label{v_HB}
V \equiv Nv_0 + N_{\rm HB}v_{\rm HB}.
\end{equation}
The tetrahedral local rearrangement does not change the average 
water-water distance, therefore the distance $r$ in
Eq.(\ref{Lennard_jones}) is not affected by the change in Eq.(\ref{v_HB}).

The enthalpy $H$ of the model is therefore
\begin{equation}\label{enthalpy}
H \equiv  \sum_{<i,j>} U(r_{ij}) + J N_{\rm HB} + J_\sigma N_{\rm coop} + PV
\end{equation}
and the volume contribution to the entropy is
\begin{equation}\label{entropy}
S_V \equiv - k_B \ln(V/v_0).
\end{equation}

The model presented here by definition does not include crystal 
phases, i.e. at any $T$ and $P$ our course grain water can be
equilibrated in its fluid state after a large enough equilibration
time. Furthermore in our coarse graining we neglect water
translational diffusion, focusing only on the local rotational
dynamics of the HBs. Formulation of the model that include translation
diffusion and crystal polymorphism can be found in Ref.~\cite{delosSantos2011} and
Ref.~\cite{oriol}, respectively, for the monolayer case. Analogous
formulations can be introduced also for the bulk model. It's worth noting, however, that the
present definition includes structural and densities heterogeneities
associated to the HB network dynamics.

\section{Simulation method}
We perform Monte Carlo (MC) simulations at constant $N=8000$, constant
$P$ and constant $T$ in a
cubic (variable) volume $V$  with periodic boundary conditions along
two axis \footnote{We adopt two-axis periodic boundary conditions in
  such a way to be able to 
    perform consistent checks with previous results for the monolayer model.}.
We update the HB network with the Wolff Monte Carlo (MC) cluster algorithm adapted
to the present model, as explained in Ref.~\cite{Mazza2009}.

We calculate the equation of state along isobars in
the range of $Pv_0/(4\epsilon) \in [-0.5, 0.95]$ separated by
intervals of $\Delta P \leq 0.05 (4\epsilon)/v_0$. The range of
temperatures is
$T k_B/(4\epsilon) \in [0.02 , 1.0]$ with $\Delta T \geq 0.01 (4\epsilon) / k_B$
if $T  k_B/(4\epsilon) \in (0.1,1.0]$, and
$\Delta T \geq 0.001 (4\epsilon) / k_B$,
if $T  k_B/(4\epsilon) \in [0.02, 0.1]$. We calculate each isobar by
sequential annealing starting  
at $T=7 (4\epsilon)/k_B$
and letting the system
equilibrate, as explained in details in Appendix \ref{app_gas-liq}.
For each state point we average over 
$10^6$ MC steps after equilibration, with a number of independent
configurations between $10^2$ -- $10^3$ depending on the  state point.

\begin{figure}[top]
\includegraphics[scale=0.3]{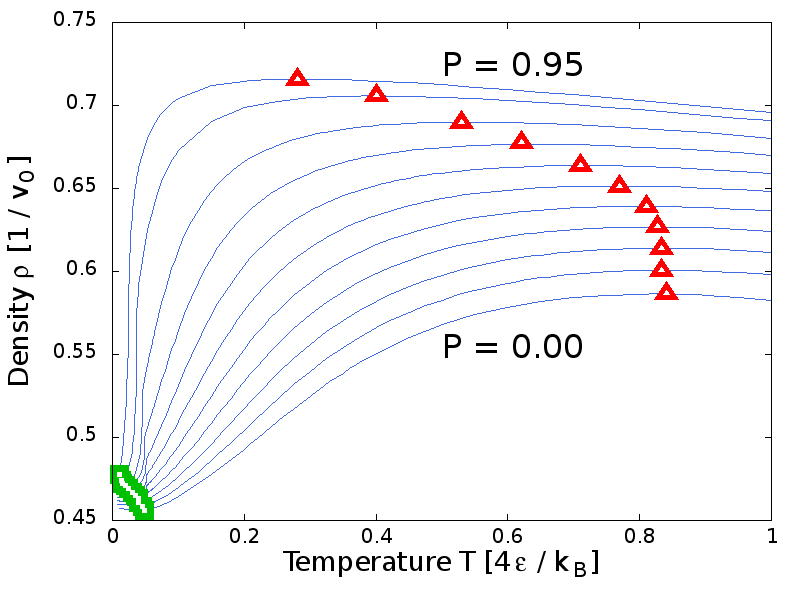}
\caption{\label{density} Density along isobars for 
  (from bottom to top) $P v_0/(4\epsilon) = 0.00$, 0.10, 0.20, 0.30, 0.40,
  0.50, 0.60, 0.70, 0.80, 0.90, 0.95.
  At each $P$ we find a temperature of maximum density, TMD (red triangles)
and  a temperature of
minimum density, TminD (green squares).
For sake of clarity we show the interpolated equation of state and do
not show the calculated state points.}
\end{figure}

\section{Results and discussion}

\begin{figure}
    a) \includegraphics[scale=0.28]{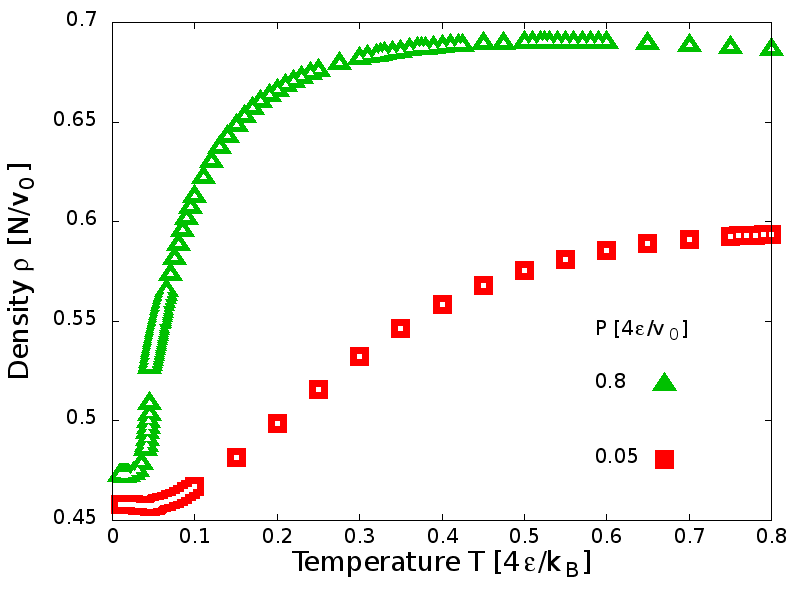}
    b) \includegraphics[scale=0.28]{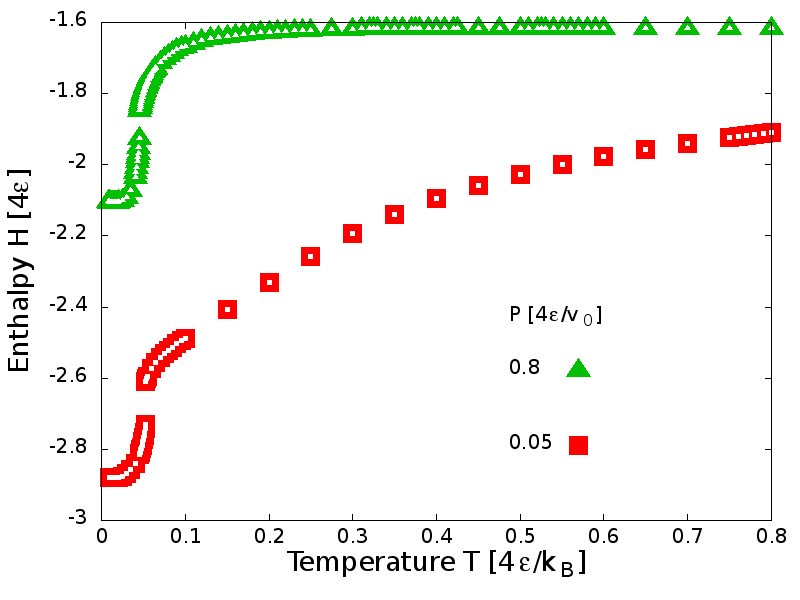}
\caption{\label{transition} a)  Isobaric variation of density  and b)
  enthalpy as a function of temperature for high pressure ($Pv_0/(4\epsilon) =0.80$, green triangles)
and   low pressure ($Pv_0/(4\epsilon) =0.05$, red squares).} 
\end{figure}

At temperatures below the gas-liquid phase transition (Appendix \ref{app_gas-liq}), we find along isobars
a temperature of maximum density (TMD) (Fig. \ref{density}) as in
water. By further decreasing $T$, we observe a sharp decrease of 
the isobaric density $\rho \equiv N/V$ 
and  a temperature of minimum
density TminD (Fig. \ref{density}). This behavior is
consistent with the 
LLPT between a high density
liquid (HDL) and a low density liquid (LDL) as postulated for 
supercooled liquid water \cite{Poole92}. However, a similar behavior,
but without any discontinuity, is predicted also by the ``singularity
free'' scenario \cite{StanleyJCP1980Vol73, Sastry1996,
  StokelyPNAS2010}. We therefore analyze the enthalpy behavior in detail.

We find that  $H$  follows the density but with sharper changes at
any $P$ (Fig. \ref{transition}).
At high $P$ both density and  enthalpy display a seemingly
discontinuous change, while for low $P$ the variation in density becomes much
smother than 
the enthalpy change.

\begin{figure}[t]
\begin{center}
a) \includegraphics[scale=0.28]{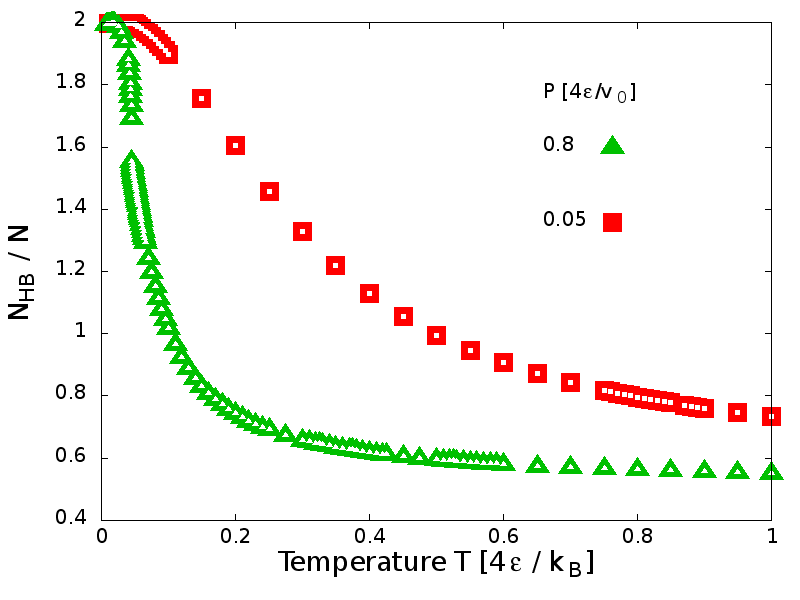}
b) \includegraphics[scale=0.28]{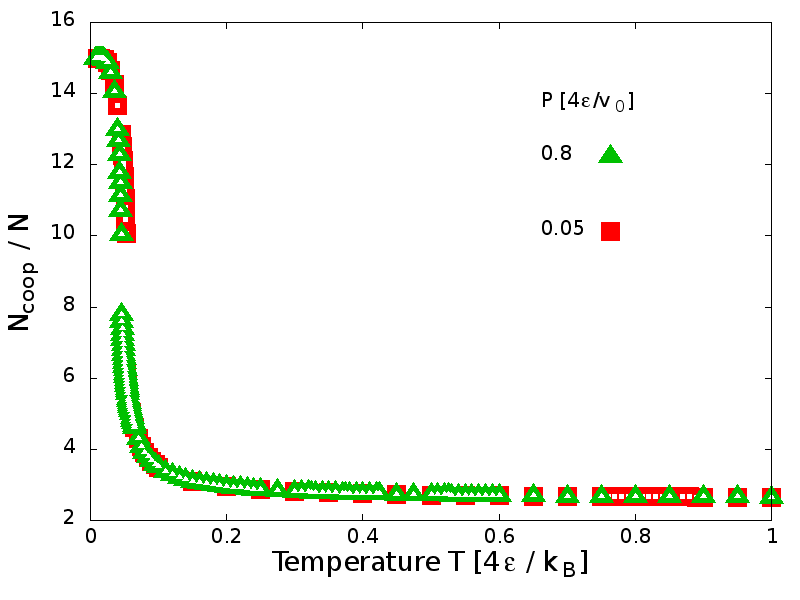}
\caption{\label{HB_transition} a) Isobaric variation of the number of
  HB   per molecule $N_{\rm HB}/N$
  and b) the number of cooperative bonds per molecule
  $N_{\rm coop} / N$.
  While $N_{\rm HB}$ strongly depend on $P$,
  $N_{\rm coop}$ is almost $P$-independent. }
\end{center}
\end{figure}

We understand the behavior of $\rho$ as
 a consequence of its dependence on
$N_{\rm  HB}$ from Eq.(\ref{v_HB}).  
A direct calculation shows
that by decreasing $T$ the model displays a rapid increase of $N_{\rm HB}$
at high $P$, while the increase is  progressive at  low $P$
(Fig. \ref{HB_transition}a).
In particular, we find that $N_{\rm HB}$ saturates at low $T$ to 
two HBs per molecule, corresponding to the case where every water
molecule is involved in four HBs. At high $T$ the number
$N_{\rm  HB}/N$ is $\simeq 1/2$ at high $P$
and is $\simeq 0.7$ at low $P$,  with each molecule forming on average one HB or
less than two, respectively.

Nevertheless, the changes of $N_{\rm HB}$ are never discontinuous as
in the enthalpy, 
clearly showing that the contribution to $H$ coming from the other
terms in Eq.(\ref{enthalpy}) are relevant.
The explicit calculation of these terms shows that the dominant
contribution comes from the behavior of
$N_{\rm coop}$ (Fig. \ref{HB_transition}b).
We find that $N_{\rm coop}$ has a sharp, but continuous, increase
at any $P$ within the range investigated. Furthermore, at variance
with what observed for  $N_{\rm  HB}$, the temperature of the
largest increase of $N_{\rm coop}$
is almost independent on $P$. This temperature
coincides with the largest variation of $H$ at any $P$ and with the
large change of $\rho$ at high $P$.
%
%
Therefore, the largest variation
of $H$ is associated with the cooperative contribution 
that, in turns, is a consequence of a large structural rearrangement
of the HBs toward a more tetrahedral configuration.
However, at low $P$ this reorganization of the HBs occurs when  
the number $N_{\rm HB}$ of HBs is almost saturated, implying only a
minor change in $\rho$. On the other hand, at high $P$ the
restructuring of the HBs occurs at the same time as the formation of a
large amount of them, up to saturation. Therefore,  the
effect on the density is  large and 
collective, as expected at a critical phase transition.  

A further way to clarify if the observed thermodynamic behavior is
consistent with the occurrence of a LLPT ending in a critical point is
to calculate the response functions $C_P$, $K_T$ and
$\alpha_P$ and to study if they diverge at the hypothesized LLCP. 
At equilibrium all these quantities correspond to thermodynamic
fluctuations. Therefore, we calculate
the specific heat along isobars both by numerical derivative of $H$, 
$C_P \equiv (\partial \langle H \rangle/\partial T)_P$, and by
the fluctuation-dissipation theorem
$C_P = \langle H^2 \rangle / k_BT^2$.
In this way we reach a better estimate of $C_P$ and at the same time, 
by verifying the validity of fluctuation-dissipation theorem, 
we guarantee  that the system is equilibrated (Appendix
\ref{app_errorcp}).

\begin{figure*}
\begin{center}
  a) \includegraphics[scale=0.3]{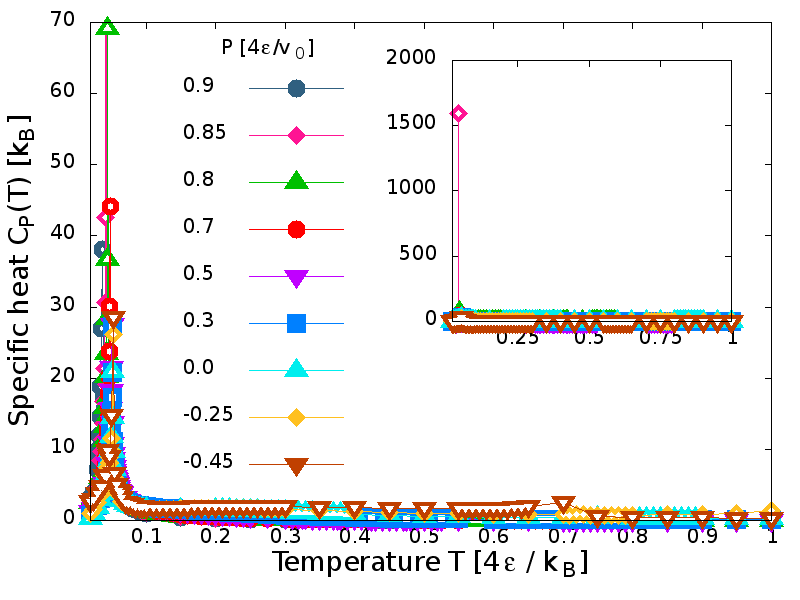}
  b) \includegraphics[scale=0.3]{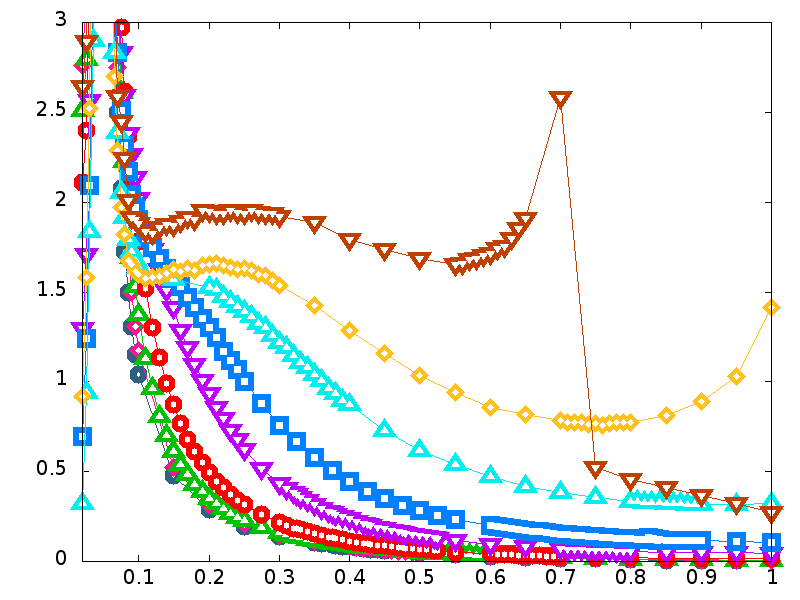}
\caption{\label{cp_maxima} Isobaric specific heat $C_P$ calculated
  as function of $T$ for different $P$.    For sake of clarity we
    show only the results from the fluctuation-dissipation theorem.
a)    We find
strong maxima at approximately constant $T$. The maxima increase for increasing 
$Pv_0/(4\epsilon)\leq 0.85$ with an apparent divergence at 
$Pv_0/(4\epsilon)= 0.85$ (inset). At higher $P$ the maxima loose
intensity and occur at lower $T$.
b) By zooming into the values of $C_P$ at
  intermediate $T$, we find also  weak maxima at negative $P$. The
  weak maxima merge into the strong maxima as
  the pressure increases. The discontinuities found at $P<0$
  and high $T$ are due to the crossing of the liquid-to-gas spinodal line.
}
\end{center}
\end{figure*}

We find strong maxima in $C_P$ 
at any $P$ and low $T$.
For $Pv_0/(4\epsilon)\leq 0.85$ the maxima occur all at the same $T$ and
increase as the pressure increases with an apparent divergence of
$C_P$ at $Pv_0/(4\epsilon)= 0.85$ (inset Fig. \ref{cp_maxima}a).
However, for
$Pv_0/(4\epsilon)> 0.85$ the maxima decrease in intensity and moves
toward lower $T$ (Fig. \ref{cp_maxima}a).
This behavior is consistent with a LLCP occurring at
$Pv_0/(4\epsilon)\simeq 0.85$ at the end of a first-order phase
transition occurring at higher $P$ along a line with a negative slope
in the $P$-$T$ thermodynamic plane, as expected in the LLCP scenario
\cite{Poole92}. 

By increasing the resolution at intermediate $T$ and exploring the
metastable region of stretched water at $P<0$, we find 
weak maxima in $C_P$ at $T$ below the liquid-to-gas spinodal line
(Fig. \ref{cp_maxima}b). For increasing $P<0$ we find that the weak
maxima occur at lower $T$ and progressively 
merge into the strong maxima.

The
origin of these maxima was studied by Mazza et al.
\cite{crossover} in the monolayer case. 
At any $P$ the weak maximum at high
temperature is produced by the energy fluctuations during the
formation of new HBs, while the strong maxima at low temperature is
produced by the effect of the cooperative reordering of the HB
network. This interpretation agrees with our results for the evolution along
isobars of $N_{\rm HB}$ and $N_{\rm coop}$
(Fig. \ref{HB_transition}).
However the weak
$C_P$ maxima for the monolayer were observed also at $P>0$ and were
slowly increasing in intensity for increasing $P$ before merging into
the strong maxima \cite{BiancoSR2014}, at variance with what we
observe now for the bulk system.

\begin{figure*}
\begin{center}
  a) \includegraphics[scale=0.3]{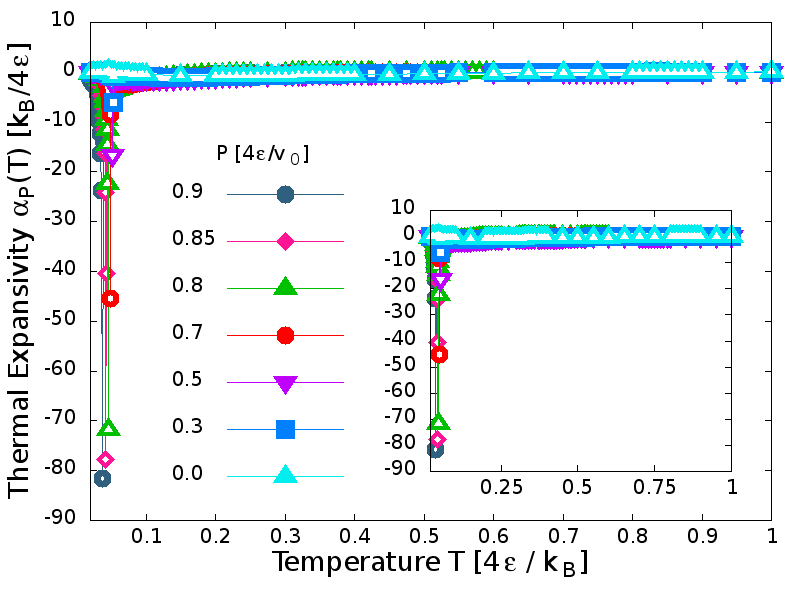}
  b) \includegraphics[scale=0.3]{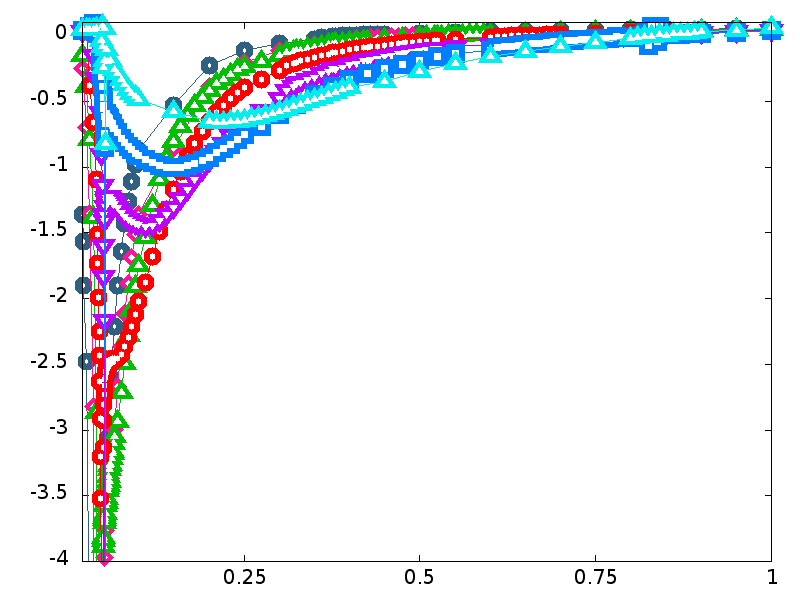}
\caption{\label{a_minima} Isobaric thermal expansivity $\alpha_P$
  calculated as function of $T$.
  a) We find strong minima at low $T$ for any $P>0$. The minima occur
  at approximately constant $T$
  with an apparent divergence at 
$Pv_0/(4\epsilon)= 0.85$ (inset). At higher $P$ the minima loose
intensity and occur at lower $T$ as for the maxima of $C_P$ (Fig.~\ref{cp_maxima}).
b) By zooming into the values of $\alpha_P$ at
  intermediate $T$, we find that the strong minima develop very
  sharp from weak minima   at higher $T$. The
  weak minima merge into the strong minima as
  the pressure increases.}
\end{center}
\end{figure*}

\begin{figure*}[t]
\begin{center}
\includegraphics[scale=0.9]{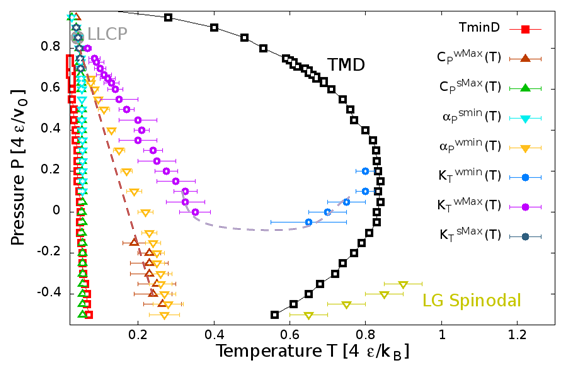}
\caption{\label{phase_diagram} The 3D many-body liquid water phase
  diagram in the $P$-$T$ plane. Below the temperature of maximum
  density (TMD) line (black squares) and
  above the liquid-to-gas (LG) spinodal (green lower triangles) at low
  $P$ and the temperature of minimum density (TminD) line (red squares),
  the loci of  
  isobaric  weak and strong extrema of, respectively,
  specific heat   $C_P^{wMax}$ (red upper triangles with dashed lines),
  $C_P^{sMax}$ (green upper triangles), thermal expansivity 
  $\alpha_P^{wmin}$ (yellow lower triangles),
  $\alpha_P^{smin}$ (turquoise lower triangles), and isothermal compressibility 
  $K_T^{wMax}$ (purple circles),
  $K_T^{sMax}$ (green circles) cross all at
  $Pv_0/(4\epsilon)\simeq 0.85$, where we 
  estimate the liquid-liquid critical point (LLCP, gray circle).
  The locus of  $K_T^{wMax}$ converges toward the  locus of isobaric  minima
  $K_T^{wmin}$ (blue circles). Dashed lines are guides for the eyes.}
\end{center}
\end{figure*}

Next,
we calculate the thermal expansivity $\alpha_P \equiv
(1 / \langle V \rangle)(\partial \langle V \rangle / \partial T)_P$
along isobars and find extrema (minima) whose behavior is similar to
those for $C_P$ (Fig.~\ref{a_minima}). Specifically, we find 
strong minima at low $T$ and weak mimima at intermediate $T$ that
merge to the strong extrema for increasing $P$ before reaching an
apparent divergence for $Pv_0/(4\epsilon)= 0.85$, consistent with the
behavior observed for $C_P$. The main difference with the results for
$C_P$ is that the strong extrema of $\alpha_P$ are very sharp in
$T$. However, both strong and weak extrema of $\alpha_P$ occur at
approximately the same temperatures as those for $C_P$
(Fig.~\ref{phase_diagram}). Furthermore, while the weak maxima of $C_P$
are observable only for $P<0$, the weak minima of $\alpha_P$ can be
calculated for any $Pv_0/(4\epsilon)\leq 0.7$.

\begin{figure*}
\begin{center}
  a) \includegraphics[scale=0.3]{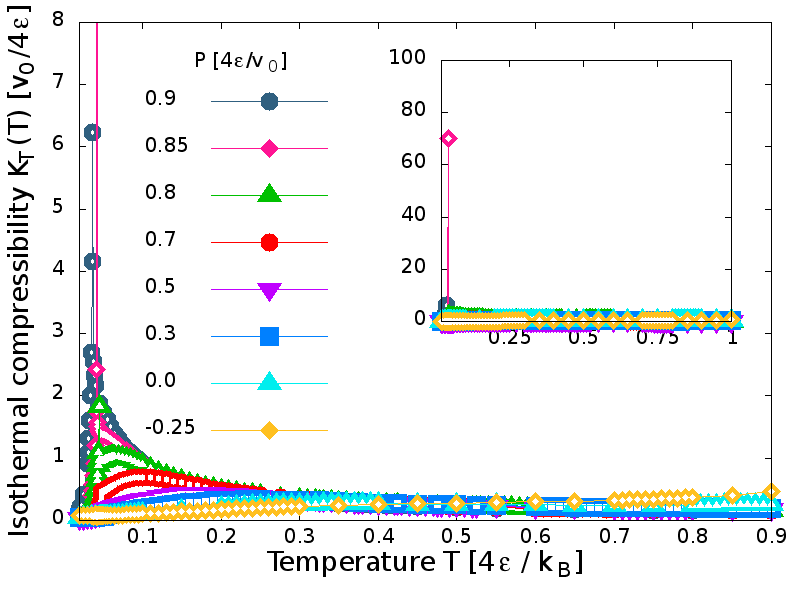}
  b) \includegraphics[scale=0.3]{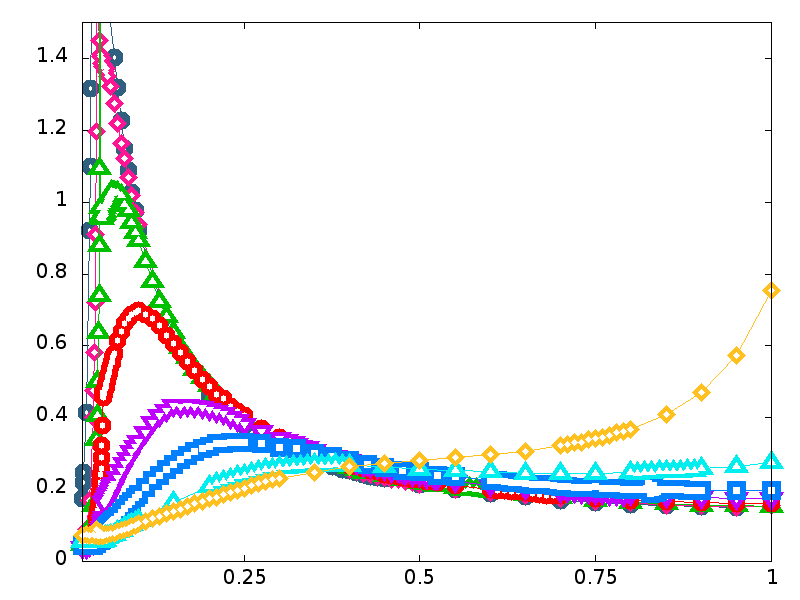}
\caption{\label{kt_maxima} Isothermal compressibility $K_T$ calculated
  along isobars from the fluctuation-dissipation theorem.
  a)    We find, at low $T$ and any $P>0$, maxima  that smoothly grows
  and occur at lower $T$ for higher $P$.
  b) A closer look at intermediate values $K_T(4\epsilon)/v_0\simeq 1$
  allows us to observe that, as for the case of $\alpha_P$
  (Fig.~\ref{a_minima}), there are sharp extrema departing from the
  smooth (weak) extrema for
  $0.75\leq Pv_0/(4\epsilon)\leq 0.90$.
  The maxima 
  occur all at approximately the same $T$ and  increase in intensity
  as $P$ approaches $Pv_0/(4\epsilon)= 0.85$ (inset in a).
The weak maxima   merge into the strong maxima at
$Pv_0/(4\epsilon)\simeq 0.85$. Finally, at high $T$ and
$0\leq Pv_0/(4\epsilon)\leq 0.2$, we
find minima in $K_T$ occurring at increasing $T$ for increasing $P$.}
\end{center}
\end{figure*}

Finally,
we calculate the isothermal compressibility $K_T$ from the 
fluctuation-dissipation theorem,
$K_T = \langle \Delta V^2 \rangle/ k_B T V$, along isobars
(Fig. \ref{kt_maxima}). 
We find strong maxima for 
$0.7<P v_0/4 \epsilon \leq 0.9$
with an apparent divergence at $Pv_0/(4\epsilon)= 0.85$ and weak
maxima for $P>0$. 
The strong maxima are sharp as for the extrema of
$\alpha_P$. Furthermore we observe that the weak maxima of $K_T$ occur
at $T$ higher than the weak extrema of $\alpha_P$ and that, approaching
$P=0$, they turn into minima. The minima occur at $T$ increasing with
$P>0$.   

In summary, the fluctuations of enthalpy associated to  $C_P$,  the
fluctuations of
volume associated to $K_T$ and the cross fluctuations of volume and
entropy associated to $\alpha_P$ increase when $T$ decreases at
constant $P$ and display two maxima: a maximum occurring at a
$P$-dependent  value of $T$, and a strong maximum occurring a very low
$T$ that is almost $P$-independent. 

The strong maxima appear at the same $T$ where $N_{\rm coop}$ has the
largest variation, i.e. where the fluctuations of  $N_{\rm coop}$ are
maxima. As a consequence, both the volume fluctuations for the
Eq.(\ref{v_HB}), and the enthalpy fluctuation for the
Eq.(\ref{enthalpy}) are maxima near the same $T$ \cite{FS2007}.
Therefore, the strong maxima are due to the cooperative rearrangement
of the HB network. 

On the other hand, the weak maxima at higher temperatures
are due to weak volume fluctuations associated to the formation of
single HBs, coinciding with the largest variation of $N_{\rm HB}$, as
already seen for the water monolayer case
\cite{MazzaPNAS2011, mazza:204502}.

The resulting phase diagram (Fig. \ref{phase_diagram}) of the 3D
many-body water closely resemble the one observed for the monolayer
case \cite{BiancoSR2014}. In the $P$-$T$ phase diagram we find the
liquid-to-gas (LG) spinodal at negative pressures defined as the
point where $K_T$ presents a huge increase and $C_P$ has a
discontinuous decrease due to the emergence of the 
gas phase.
Above the LG spinodal we find  a retracing TMD line
that converges at low $P$ toward the TminD line at low $T$, consistent with
experimental data \cite{mallamace} and atomistic simulations \cite{minima}.
Below the TMD line we find weak extrema at higher $T$ and strong
extrema at lower $T$ for the
response functions, $C_P(T)$, $K_T(T)$ and $\alpha_P(T)$.
While the loci of weak extrema are $P$-dependent and merge at high $P$,
the loci of strong extrema are $P$-independent
and overlap.
The locus of weak maxima of $K_T$ converges at lower $P$ toward the
locus of minima of $K_T$ as observed in atomistic simulations
\cite{minima}. 
Furthermore, as can be demonstrated by thermodynamic argument
\cite{minima}, the 
locus of minima of $K_T$ crosses the turning point of the TMD line,
showing that our results are thermodynamically consistent \cite{BiancoSR2014}.

All the loci of strong and weak extrema converge toward the same
high-$P$ region where they all display a large increase near 
$Pv_0/(4\epsilon)\simeq 0.85$ where they all seem to diverge. 
As demonstrated by finite size scaling for the case of a
monolayer  \cite{BiancoSR2014}, this finding is consistent with the
occurrence of a LLPT ending in a liquid-liquid critical point.

\section{Conclusions}


Our analysis shows that the bulk many-body
water model reproduces the fluid properties of
water in a wide range of pressures and temperatures.
Below the liquid
gas phase transition and above the liquid-to-gas spinodal in the
stretched liquid state at $P<0$, we find the TMD line and, at much
lower $T$, the TminD line. The first at atmospheric pressure,
corresponding to $P\simeq 0$ in our model, occurs at
277~K, while the second is estimated at a $T= 203 \pm 5$~K
\cite{Mallamace2007}. Therefore, the phase diagram we present here
goes from the gas to the deep supercooled region of liquid water.  

Below the TMD line, we find $P$-dependent extrema of the response
functions $C_P$, $\alpha_P$ and $K_T$. While  $\alpha_P$ has extrema
at all the investigated $P$,  $C_P$ displays  $P$-dependent maxima only
at negative $P$, while $K_T$ only at $P>0$. In particular, the maxima
of $K_T$ converge toward minima for $P$ approaching 0 and these minima
cross the turning point of the TMD line as expected
\cite{minima}, showing that our results are thermodynamically
consistent and similar to those of 
atomistic
models for water \cite{minima, Gonzales2016}.

However, our model allows us to simulate in an efficient way
supercooled liquid water even at
temperatures below the TminD line, an extremely demanding task for
atomistic models as a consequence of the glassy slowing down of the
dynamics \cite{kesselring2012, kesselringJCP2013}.
Thanks to this peculiar property we find that at
$Tk_B/4\epsilon \simeq 0.035$,
below the loci of
the $P$-dependent maxima, all the response functions have sharp maxima
that have almost no $P$ dependence up to
$Pv_0/(4\epsilon)\simeq 0.85$.
We therefore call these maxima strong and call weak the $P$-dependent
maxima. We find that the weak maxima merge with the strong maxima
for $Pv_0/(4\epsilon)\leq 0.85$.

It is interesting to observe that weak and strong maxima have been
found also in 
the monolayer many-body water model
\cite{MazzaPNAS2011,  mazza:204502,  BMSBSF2013,  BiancoSR2014} but
with an important difference. 
For the monolayer case the strong maxima of $C_P$ are observed also
for $P>0$, while for the bulk case they are overshadowed  by the
strong maxima  at $P>0$.

This difference between monolayer and bulk water is consistent with
the fact that  in the
latter the total number of accessible states is much larger. In the
case of the monolayer  each molecule has four bonding
variables with $6^4$ accessible configurations \cite{delosSantos2011}. Instead,  
for the bulk each molecule has six bonding
variables with $6^6$ accessible configurations.
Therefore, the entropy loss due to the  formation of the HB network is
much greater for the bulk with respect to the monolayer, consistent
with a $C_P$ with broader maxima that dominate over the weak maxima.

It is worth noticing that although our prediction of strong and weak maxima
for the response functions has not been directly
tested in atomistic models, it is perfectly consistent, qualitatively, with the
results of atomistic simulations and at the same time with the
experimental results 
\cite{BMSBSF2013}.
Furthermore, several atomistic models give hints of more than one
maxima in $C_P$, showing  the line of $P$-dependent $C_P$ maxima
converging toward the line of $C_P$ minima when crossing the 
point where the TMD line meets the TminD line
\cite{minima,  Gonzales2016}. This convergence of loci of $C_P$
maxima, TMD, TminD and $C_P$ minima is consistent with the results for
the water monolayer \cite{BiancoSR2014} and possibly also with the
present results for bulk water.

We find that all the maxima of response functions merge and diverge at 
$Pv_0/(4\epsilon)\simeq 0.85$ and $Tk_B/4\epsilon \simeq
0.035$. Therefore we estimate the occurrence of a LLCP at these
approximate values.
A finite-size analysis, as the one performed for the monolayer case
\cite{BiancoSR2014}, would be necessary to estimate with larger 
confidence the LLCP. This analysis could also allows us to study the
universality class of the LLCP in bulk. However, such analysis is
beyond the scope of the present work.

For $Pv_0/(4\epsilon)> 0.85$ we find that the
maxima of the response functions decrease and occur at lower $T$,
consistent with a LLPT with negative slope in the $P$-$T$ plane and
ending in the LLCP, as found for the water monolayer
\cite{BiancoSR2014}.

In conclusion, our results for the bulk many-body water model are
consistent  with the LLCP scenario for supercooled liquid water
\cite{Poole92}. Furthermore, we find properties consistent  with those 
demonstrated in a rigorous way for the many-body
water monolayer, including the occurrence of strong and weak maxima
for the response functions \cite{BiancoSR2014}.
Therefore, we can argue that these properties do not
originate in the low-dimensionality of  
the monolayer system, but they are an intrinsic property of the model
due to the cooperative contribution to the HB interactions.
Because the cooperativity in water is a necessary implication of its
peculiar properties \cite{Barnes1979, Guevara-Vela:2016aa}, we argue
that our results are relevant for real  
water.

\section{Acknowledgments}
 V.B. and G.F. acknowledge support from the Austrian Science Fund
(FWF) project P 26253-N27 and the Spanish MINECO grant FIS2012-31025
 and FIS2015-66879-C2-2-P, respectively.


\appendix

\section{Gas-liquid phase transition}\label{app_gas-liq}
\begin{figure}[t]
\begin{center}
\includegraphics[scale=0.3]{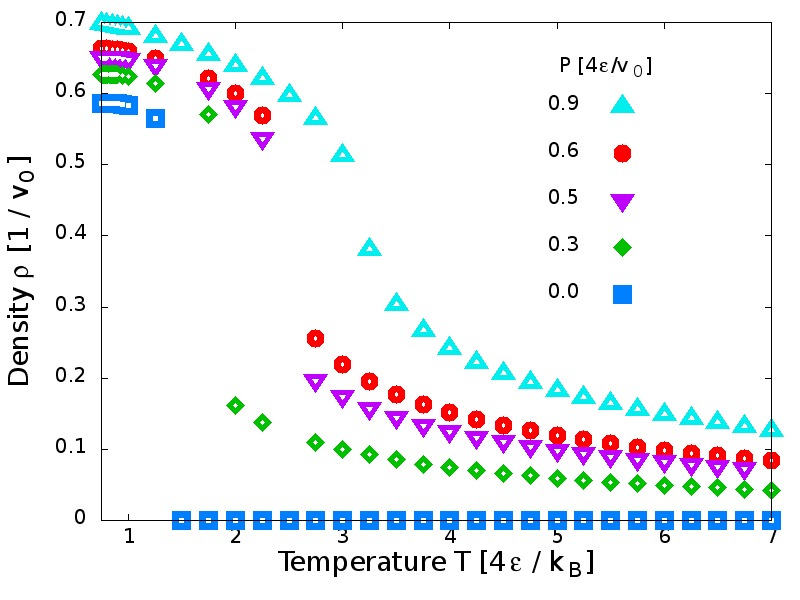}
\caption{\label{gas_density} Isobaric variation of the density at high temperatures. There is an abrupt change in the density as the system undergoes the gas-liquid phase transition}
\end{center}
\end{figure}

We calculate  isobars by applying an annealing protocol.
We first
simulate
a random configuration at high
temperature. Next we decrease the temperature and
simulate starting from the equilibrated
configuration of the previous step.
Before reaching the liquid phase, water undergoes a gas-liquid phase
transition (Fig. \ref{gas_density}) with an abrupt increase of
density in the transition from gas to liquid phase.

\section{Error calculation in the specific heat}\label{app_errorcp}
We made two independent calculations for the specific heat, one 
using the fluctuation-dissipation 
(FD) theorem and the other, using its definition as the temperature
derivative of the enthalpy (TD). According to Statistical Physics,
both results are equal under thermodynamic equilibrium conditions. For
this reason, a good test to check whether the system is or is not
equilibrated is to compare both results. If the system is
well-equilibrated, they must overlap within their error bars. 

\begin{figure}
\begin{center}
\includegraphics[scale=0.3]{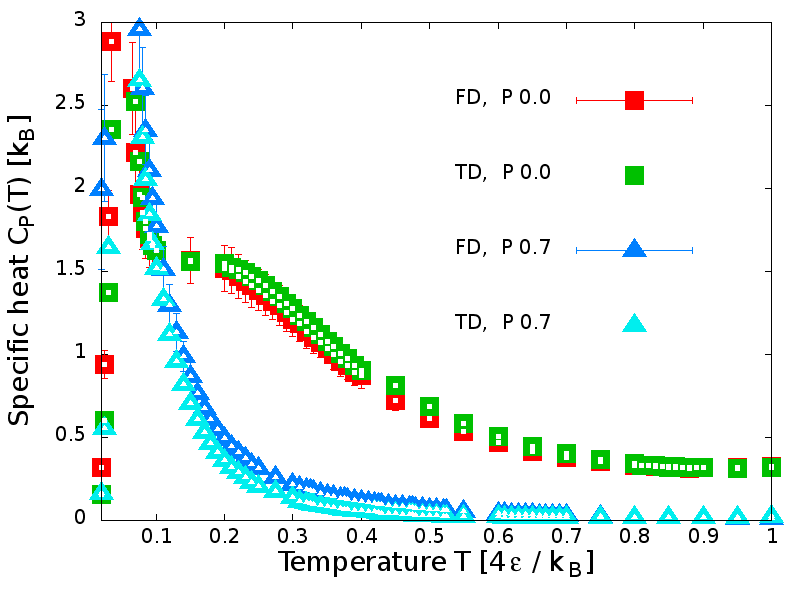}
\caption{\label{cp_error} Isobaric specific heat for two cases of high and low pressure. The resulting $C_P$ obtained from TD fits within the errorbar of FD}
\end{center}
\end{figure}

We calculate average enthalpy $\langle H \rangle$ and fluctuations
$\langle \Delta H^2 \rangle$ from the simulations and estimate 
the error using the Jackknife method.
The method allows us also to estimate 
the MC correlation time as 
$\tau^{MC} \sim 256$ MC Steps.
As a consequence we estimate that our averages are calculated 
over 
$\simeq 400$ independent simulations (Fig. \ref{cp_error}).

On the one hand, the FD formula for the specific heat $C_P$ and its error $\delta C_P$ are
\begin{equation}\label{FD_cp}
C_P^{FD} = \frac{<\Delta H^2>}{k_B T^2}
\end{equation}
\begin{equation}\label{FD_error}
\delta C_P^{FD} = \left| \frac{\partial C_P}{\partial \langle \Delta H^2 \rangle}\right| \delta \Delta H^2 = \frac{\delta \Delta H^2}{k_B T^2}
\end{equation}
On the other hand, the TD formula is
\begin{equation}\label{TD_cp}
C_P^{TD} = \frac{\langle H (T) \rangle - \langle H(T-\Delta T) \rangle}{\Delta T}
\end{equation}
where we have used the numerical finite differences method. The error for $C_P$ in the TD method is
\begin{eqnarray}
\delta C_P^{TD} = \frac{\delta \langle H(T) \rangle + \delta \langle
  H(T - \Delta T)\rangle}{\Delta T}.
\end{eqnarray}

\providecommand{\noopsort}[1]{}\providecommand{\singleletter}[1]{#1}%

\end{document}